\documentclass[reprint,a4paper,showpacs,showkeys]{revtex4-1}
\usepackage[utf8]{inputenc}
\usepackage[english]{babel}
\usepackage{graphicx}
\usepackage{dcolumn}
\usepackage{textcomp}
\newcommand{\TeOii}{\mbox{$\mathrm{TeO}_2$}}
\newcommand{\NiO}{\mbox{$\mathrm{NiO}$}}
\newcommand{\WOiii}{\mbox{$\mathrm{WO}_3$}}
\newcommand{\WOiiiTeOiix}[2]{%
\mbox{$#1\mathrm{WO}_3\textrm{--}#2\mathrm{TeO}_2$}}
\newcommand{\WOiiiTeOiixNi}[2]{%
\mbox{$#1\mathrm{WO}_3\textrm{--}#2\mathrm{TeO}_2\textrm{:}\mathrm{Ni}^{2+}$}}
\newcommand{\Niiip}{\mbox{$\mathrm{Ni}^{2+}$}}
\newcommand{\TMxp}[2]{\mbox{$\mathrm{#1}^{#2+}$}}
\newcommand{\NiOTe}{%
\mbox{$\mathrm{Ni}\!\!\relbar\!\!\mathrm{O}\!\!\relbar\!\!\mathrm{Te}$}}
\newcommand{\Symm}[2]{\mbox{$\mathrm{#1}_{#2}$}}
\newcommand{\Term}[6]{%
\mbox{${}^{#3}\mathrm{#1}_{#2}\left({}^{#6}\mathrm{#4}_{#5}\right)$}}
\newcommand{\dn}[1]{\mbox{$d^{\,#1}$}}
\newcommand{\QE}{\mbox{Quantum-Espresso}}
\newcommand{\AOMX}{\mbox{AOMX}}
\newcommand{\mkm}{\textmu{}m}
\newcommand{\cmminusi}{$\textrm{cm}^{-1}$}
\newcommand{\dbkmppm}{%
$\textrm{dB~km}^{-1}\left(\textrm{wt.~ppm~Ni}\right)^{-1}$}
\newcommand{\wtNi}{wt.\%~Ni}
\begin{document}
\title{Optical absorption and structure of impurity Ni$^{2+}$ center
in tungstate-tellurite glass}
\author{V.G.Plotnichenko}
\email[E-mail:~~]{victor@fo.gpi.ac.ru}
\author{V.O.Sokolov}
\email[E-mail:~~]{sokolov@fo.gpi.ac.ru}
\affiliation{Fiber~Optics~Research~Center of the~Russian~Academy~of~Sciences \\
38~Vavilov~Street, Moscow 119333, Russia}
\author{G.E.Snopatin}
\author{M.F.Churbanov}
\affiliation{Institute of~Chemistry of~High-Purity~Substances
of the~Russian~Academy~of~Sciences \\
49~Tropinin~Street, Nizhny~Novgorod 603600, Russia}

\begin{abstract}
Absorption spectra of \Niiip{} ions in \WOiiiTeOiix{22}{78}{}
tungstate-tellurite glass were studied and \Niiip{} extinction coefficient
spectral dependence was derived in the 450 -- 2700~nm wavelength range.
Computer modeling of the glass structure proved \Niiip{} ions to be in
trigonal-distorted octahedral environment in the tungstate-tellurite glass.
Tanabe-Sugano diagram for such an environment was calculated and good
description of the observed spectrum of \Niiip{} ion was obtained. Basing on
both absorption spectral range width and the extinction coefficient, nickel
should be considered among the most strongly absorbing impurities in the
tellurite glasses.
\end{abstract}

\pacs{71.55.-i, 71.70.Ch, 78.20.Bh, 78.20.Ci, 78.40.Pg }
\keywords{%
tellurite glasses;
defect centers;
glasses containing transition metal ions;
optical spectroscopy;
computer simulation
}

\maketitle

\section{Introdiction}
\label{Introdiction}
Tellurite glasses are known to have wide transmission range (0.35 -- 6.0~\mkm),
high linear and nonlinear refractive indices and potentially low optical losses
in the near- and mid-IR ranges. Because of this tellurite glasses are of
considerable interest as fiber optics materials. Glasses intended for optical
fibers manufacturing should contain low amount of impurities responsible for
optical loss. $3d$ transition metals ions (V, Cr, Mn, Fe, Co, Ni and Cu) giving
rise to intense absorption in the visible and near infrared region represent
one of the main groups of the limiting impurities.

In the literature there are no quantitative data on the influence of $3d$
transition metals impurities on optical absorption in the transmission spectral
range of tellurite glasses. On the other hand, absorption spectra of $3d$
transition metals ions in zirconium fluoride- and silica-based glasses are
studied to a considerable extent \cite{Newns73, Ohishi83, Day90, Schultz74,
France85}. In zirconium fluoride-based glasses the optical absorption in the 1
-- 2~\mkm{} range is limited by \TMxp{Co}{2}, \TMxp{Fe}{2}{} and \Niiip{}
impurities with extinction coefficient 100 -- 500~dB~km$^{-1}$~ppm$^{-1}$
\cite{Day90}. Cr, Co, Fe and Cu impurities with extinction coefficient 300 --
700~dB~km$^{-1}$~ppm$^{-1}$ in the silica transparency range, 1.3 -- 1.6~\mkm,
are known to be the most limiting impurities in silica fibers \cite{Schultz74}.

Extinction coefficient relates optical absorption intensity to impurities
content. Hence knowledge of the extinction coefficient spectral dependence is
needed, on the one hand, to control undesirable impurities content in optical
materials using the materials transmission spectra, to develop justified
requirements for the acceptable content of impurities and to adjust technology
to improve the optical parameters of materials; on the other hand, to control
the content of dopants explicitly added to the material to optimize the
characteristics of the developed lasers, optical amplifiers and converters.

The purpose of the present work is to study the transmission spectra of
nickel-doped tungstate-tellurite glasses, \WOiiiTeOiixNi{}{}, to determine the
spectral dependence of \Niiip{} extinction coefficient in the 450 -- 2700~nm
wavelength range and to relate absorption spectra of \Niiip{} ions in the
tungstate-tellurite glass with structure of their environment.
Tungstate-tellurite glasses is known to be one of the most
crystallization-resistant binary tellurite glasses. Nickel as a dopant is chosen
for the investigation on account of its abundance in oxides used as raw
materials and of the only valence state, $+2$, of nickel in oxides. The latter
is of importance in view of oxidizing nature of \TeOii{} as macro component of
the glass under consideration.

\section{Experimental}
\label{Experimental}
\subsection{Preparation of Ni$^{\;2+}$-doped tungstate-tellurite glasses}
Tungstate-tellurite glasses were prepared by cooling molten
\WOiiiTeOiix{22}{78}{} high-purity oxides mixture \cite{Churbanov08}. \TeOii{}
and \WOiii{} oxides with nickel impurity content as low as $1.0 \cdot
10^{-4}$~\wtNi{} were used as raw components. Mixture of powdered raw
components were pre-dried to remove water and then melted in platinum crucible
heated by high-frequency inductor placed in oxygen atmosphere with 0.8~ppm
humidity. The melt was homogenized for 1~hour at a temperature of
$800^\circ\textrm{C}$ (monitored by an optical pyrometer).

After homogenizing melting, part of the melt was solidified into
\WOiiiTeOiix{22}{78}{} glass for later use to dilute nickel-doped glass melts.
Nickel oxide was added to the rest of the melt to obtain doped glass containing
1~wt.\%~NiO. Once the nickel-doped melt was homogenized, part of it was
vitrified and the previously prepared \WOiiiTeOiix{22}{78}{} diluent glass was
added to the rest of the melt in the crucible. Repeating the process several
times, \WOiiiTeOiixNi{22}{78}{} glasses containing from 0.786 up to $9.4 \cdot
10^{-3}$~\wtNi{} were obtained. Nickel oxide concentration in the solidified
glass was calculated from material balance equation.

\WOiiiTeOiixNi{22}{78}{} glasses samples with thickness from 0.5 to 50~mm with
optically polished faces were prepared. Transmission spectra of these samples in
visible and infrared spectral ranges were measured using Perkin Elmer Lambda~900
spectrophotometer (in the 400 -- 3000~nm range) and Bruker IFS~113v Fourier
spectrometer (in the 1000 -- 10000~nm range) with spectral resolution better
than 4~nm.
\begin{figure}
\includegraphics[scale=0.50,bb=60 300 550 740]{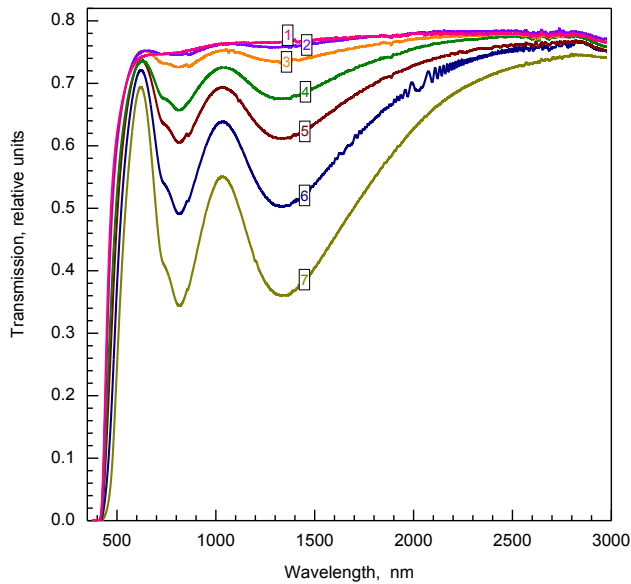}
\caption{Transmission spectra of \WOiiiTeOiixNi{22}{78}{} tellurite glasses
0.5~mm-thick samples with different nickel oxide content (\fbox{1}~0.012,
\fbox{2}~0.029, \fbox{3}~0.068, \fbox{4}~0.16, \fbox{5}~0.27,
\fbox{6}~0.51, \fbox{7}~1.01~wt.\%~NiO)}
\label{fig:1}
\end{figure}
\begin{figure}
\includegraphics[scale=0.50,bb=60 300 550 740]{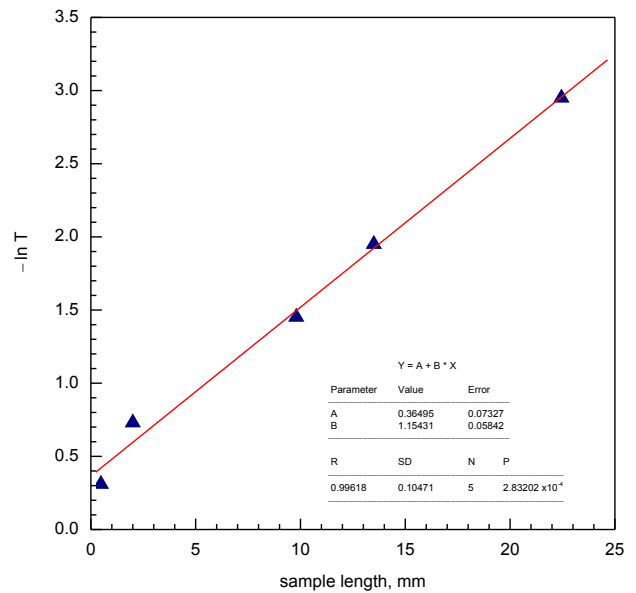}
\caption{Dependence of the radiation attenuation in the maximum of the 1320~nm
absorption band in \WOiiiTeOiixNi{22}{78}{} tellurite glasses containing
0.053~\wtNi{} on the sample length and its linear approximation}
\label{fig:2}
\end{figure}

\subsection{Measurements}
Fig.~\ref{fig:1} shows transmission spectra of 0.5~mm-thick samples containing
from 0.009 to 0.786~\wtNi. As viewed in Fig.~\ref{fig:1}, in the absorption
spectrum of \WOiiiTeOiixNi{22}{78}{} glass there are two broad bands with maxima
near 810 and 1320~nm and a band with maximum at wavelength $< 450$~nm falling
within the short-wavelength edge of \WOiiiTeOiixNi{22}{78}{} glass transparency
range. The absorption intensity in all the bands increases with nickel content
growth.
\begin{figure}
\includegraphics[scale=0.50,bb=60 300 550 740]{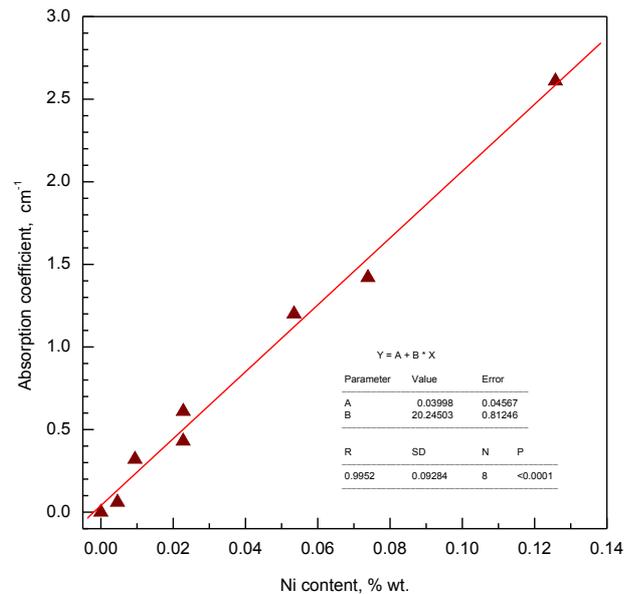}
\caption{Dependence of the bulk absorption coefficient at 1320~nm wavelength on
the nickel content in \WOiiiTeOiixNi{22}{78}{} tellurite glasses}
\label{fig:3}
\end{figure}
\begin{figure}
\includegraphics[scale=0.50,bb=60 300 550 740]{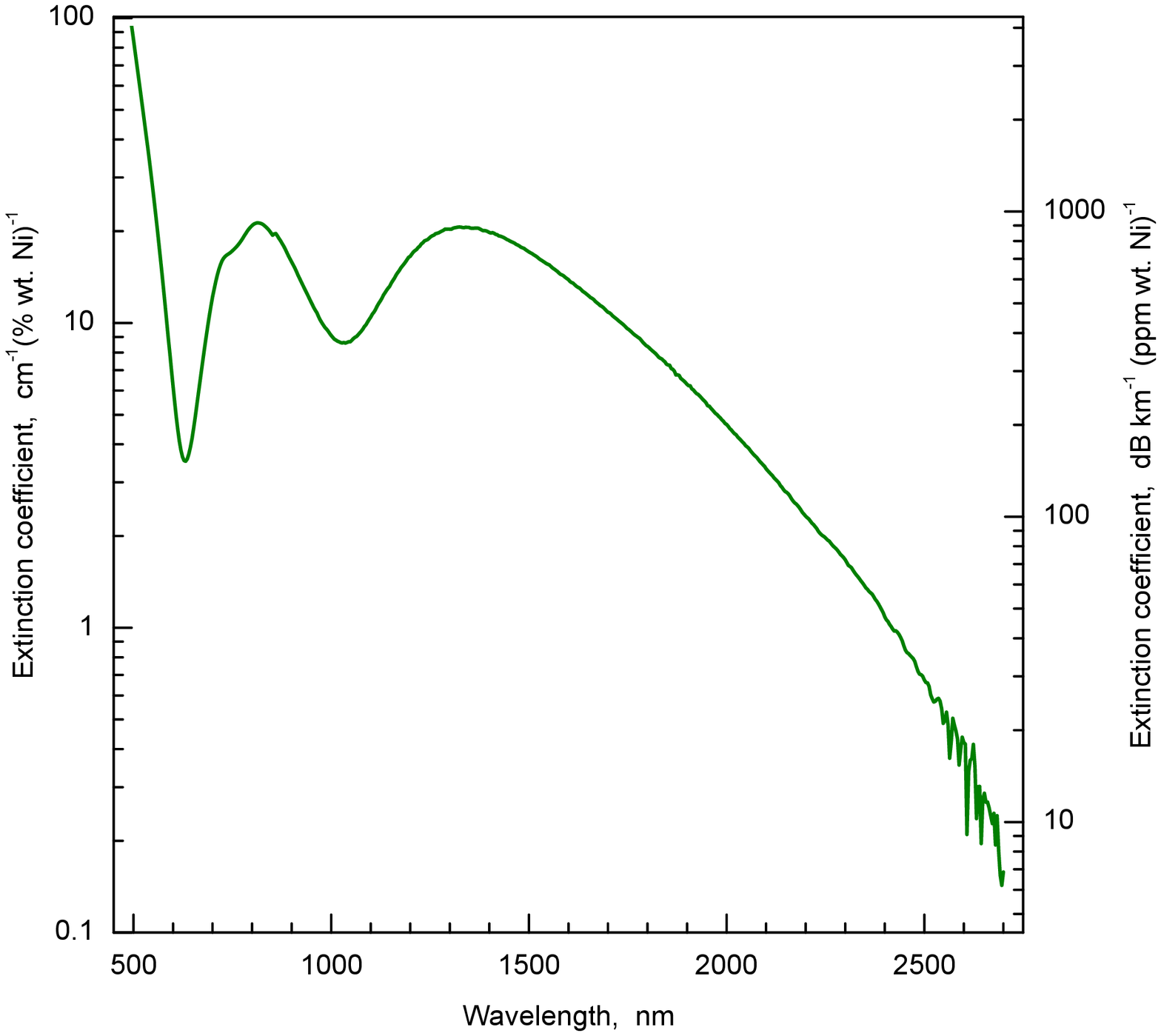}
\caption{Spectral dependence of the extinction coefficient of \Niiip{} impurity
ions in \WOiiiTeOiixNi{22}{78}{} tellurite glass}
\label{fig:4}
\end{figure}

Transmission spectra were measured for the optical path length from 0.5 to 22~mm
for samples with different nickel content. This allowed us to determine the
extinction coefficient relating absorption of \Niiip{} ion with concentration of
such ions in \WOiiiTeOiixNi{22}{78}{} tellurite glass in the 450 -- 2700~nm
spectral range. By the way of example, Fig.~\ref{fig:2} shows the light
attenuation in the peak of the 1320~nm band in glasses containing 0.053~\wtNi{}
vs. the sample thickness together with linear approximation of the experimental
points. The $A$ coefficient of the approximation is mostly due to Fresnel
reflection losses and the line slope ($B$ coefficient) represents the bulk
absorption coefficient of the bulk glass with given nickel content.

In Fig.~\ref{fig:3} are shown the experimental data and the approximation of the
bulk absorption coefficient for the 1320~nm  band for all the samples studied
with Ni content up to approximately 0.12~\wtNi. As evident from
Fig.~\ref{fig:3}, in this range of nickel content absorption in the maximum of
the 1320~nm  band is well fitted by a linear function of the absorbing \Niiip{}
ions concentration. The straight line slope representing the extinction
coefficient is equal to $20.2 \pm
0.8\textrm{~cm}^{-1}\left(\textrm{wt.\%~Ni}\right)^{-1}$ or $870 \pm
35$~\dbkmppm.

Since both the number of absorption bands and their shapes do not depend on
nickel content in the glass, it is possible to derive the spectral dependence of
the extinction coefficient shown in Fig.~\ref{fig:4}. For reference,
Table~\ref{tab:1} presents \Niiip{} impurity ions absorption at different
wavelengths in several glasses with 1~ppm Ni content. The above-obtained
\Niiip{} extinction coefficient spectral dependence allows us to estimate the
content of impurity nickel ions in \WOiiiTeOiixNi{}{}{} tellurite glasses
acceptable to get certain optical losses in glass at different wavelengths.
5Thus, optical losses as low as 100~$\textrm{dB~km}^{-1}$ in the 600 -- 2000~nm
spectral range are possible for nickel content not more than $1 \cdot
10^{-5}$~\wtNi{} or 0.1~wt.~ppm~Ni.
\begingroup
\squeezetable
\newlength{\Li}  \settowidth{\Li}{Zr-Ba-La-Al-Na-P-F}
\newlength{\Lii} \settowidth{\Lii}{paper}
\begin{table}
\caption{Absorption (dB~km$^{-1}$) caused by 1~wt.~ppm impurity \Niiip{} ions
in various glasses}
\begin{tabular}{c|D{.}{}{1}|D{.}{}{1}|D{.}{}{1}|D{.}{}{1}|c}
\hline\hline
\multicolumn{1}{c|}{} & \multicolumn{4}{c|}{} & \multicolumn{1}{c}{} \\[-1.2ex]
glass  & \multicolumn{4}{c|}{absorption band wavelength, \mkm} &
\multicolumn{1}{c}{Ref.} \\
\multicolumn{1}{c|}{} & \multicolumn{4}{c|}{} & \multicolumn{1}{c}{} \\[-1.2ex]
\cline{2-5}
\multicolumn{1}{c|}{} &&&& \multicolumn{1}{c}{} \\[-1.2ex]
       &
\multicolumn{1}{c|}{0.4} & \multicolumn{1}{c|}{1.5} &
\multicolumn{1}{c|}{2.0} & \multicolumn{1}{c|}{2.5} &        \\
\multicolumn{1}{c|}{} &&&& \multicolumn{1}{c}{} \\[-1.2ex]
\hline\hline  &&&&& \\
\parbox[c]{\Li}{\centering $\mathrm{ZrF}_4$-based \\ (Zr–Ba–La–Al–Na–Pb–F)}
       &  650.  &  200.   &  90.   &  30.   & \cite{Ohishi83} \\
&&&&& \\[-1.2ex]
\parbox[c]{\Li}{\centering $\mathrm{ZrF}_4$-based \\ (Zr–Ba–La–Al–Na–F)}
       &  360.  &  100.   &< 30.   & < 3.   & \cite{France85} \\
&&&&& \\[-1.2ex]
$\mathrm{SiO}_2$
       & 2500.  &  200.   & \multicolumn{1}{c|}{---} & \multicolumn{1}{c|}{---}
& \cite{Schultz74} \\
&&&&& \\[-1.2ex]
\parbox[c]{\Li}{\centering \WOiiiTeOiix{22}{78}}
       &>5000.  &  750.   & 200.   &  30.   &
\parbox[c]{\Lii}{this \\ paper} \\ [2.4ex]
\hline\hline
\end{tabular}
\label{tab:1}
\end{table}
\endgroup

\section{Modeling of the WO$_{3}$--TeO$_{2\,}$:Ni$^{2+}$ system}
\label{Modeling}
Modeling of the structure of impurity nickel centers in tungstate-tellurite
glass network was performed using periodic model constructed on the basis of the
supercell containing 32 \TeOii{} groups (96 atoms in total), with the initial
atoms arrangement corresponding to paratellurite lattice. Two \TeOii{} groups in
the supercell were substituted by \NiO{} and \WOiii{} ones. So the supercell
composition was $\left(\mathrm{NiO}\right) \left(\mathrm{WO}_3\right)
\left(\mathrm{TeO}_2\right)_{30}$. The described model was used to find the
equilibrium configurations of impurity nickel atoms and tungsten atoms in the
\WOiiiTeOiix{}{}{} tellurite glass network by means of ab~initio
(Car-Parrinello) molecular dynamics \cite{Car85} with final complete geometry
optimization by gradient method. All calculations were performed in the
generalized gradient approximation of density functional theory in the plane
waves basis using the \QE{} package \cite{QE}. Calculations were performed in
two approaches, using either norm-conserving Troullier-Martins pseudopotentials
\cite{Troullier91} or ultrasoft pseudopotentials \cite{Vanderbilt90, Laasonen91,
Laasonen93}. The norm-conserving Troullier-Martins pseudopotentials and
ultrasoft pseudopotentials were developed for the tellurium atoms, oxygen,
nickel and tungsten with the help of fhi98PP \cite{Fuchs99} and USPP~v.~7.3.6
\cite{Vanderbilt06} programs, respectively. $3d$, $4s$ and $4p$ shells were
taken to be the valence ones in the case of nickel atom.
\begin{figure}
\includegraphics[scale=0.31,bb=85 40 875 810]{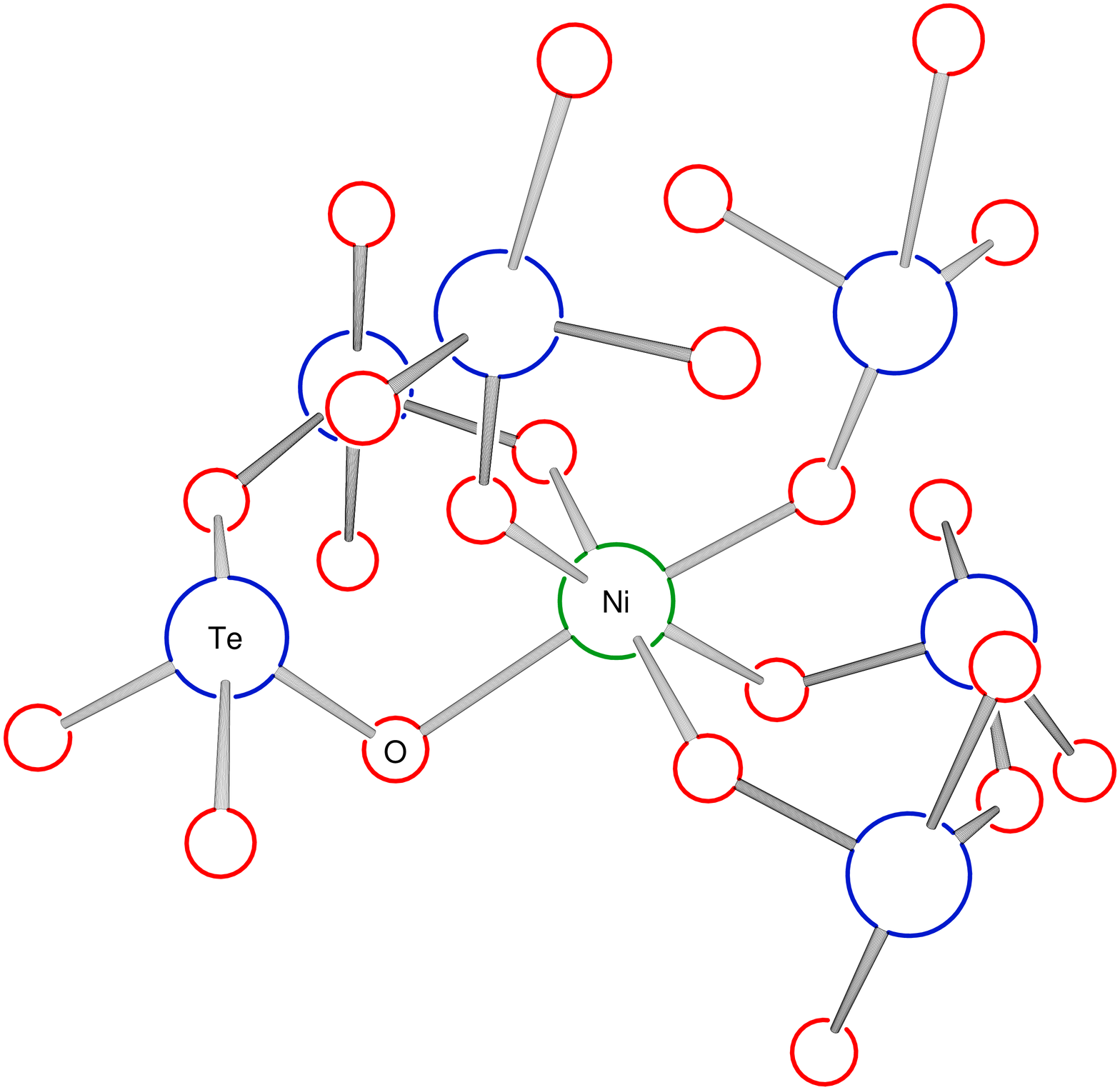}
\caption{Calculated configuration of Ni impurity atom in \WOiiiTeOiix{}{}{}
tellurite glass network}
\label{fig:5}
\end{figure}

Calculations showed that \Niiip{} ion is sixfold coordinated in the
\WOiiiTeOiix{}{}{} glass network ion, its environment being trigonally
(\Symm{D}{3}) distorted octahedron with oxygen atoms in its vertices forming
\NiOTe{} linkages (Fig.~\ref{fig:5}). It should be remarked that \Niiip{} turns
out to be surrounded by pure \TeOii{} network: no tungsten atom occurs in
three nearest coordination shells.

\begin{figure}
\includegraphics[scale=0.50,bb=60 300 550 740]{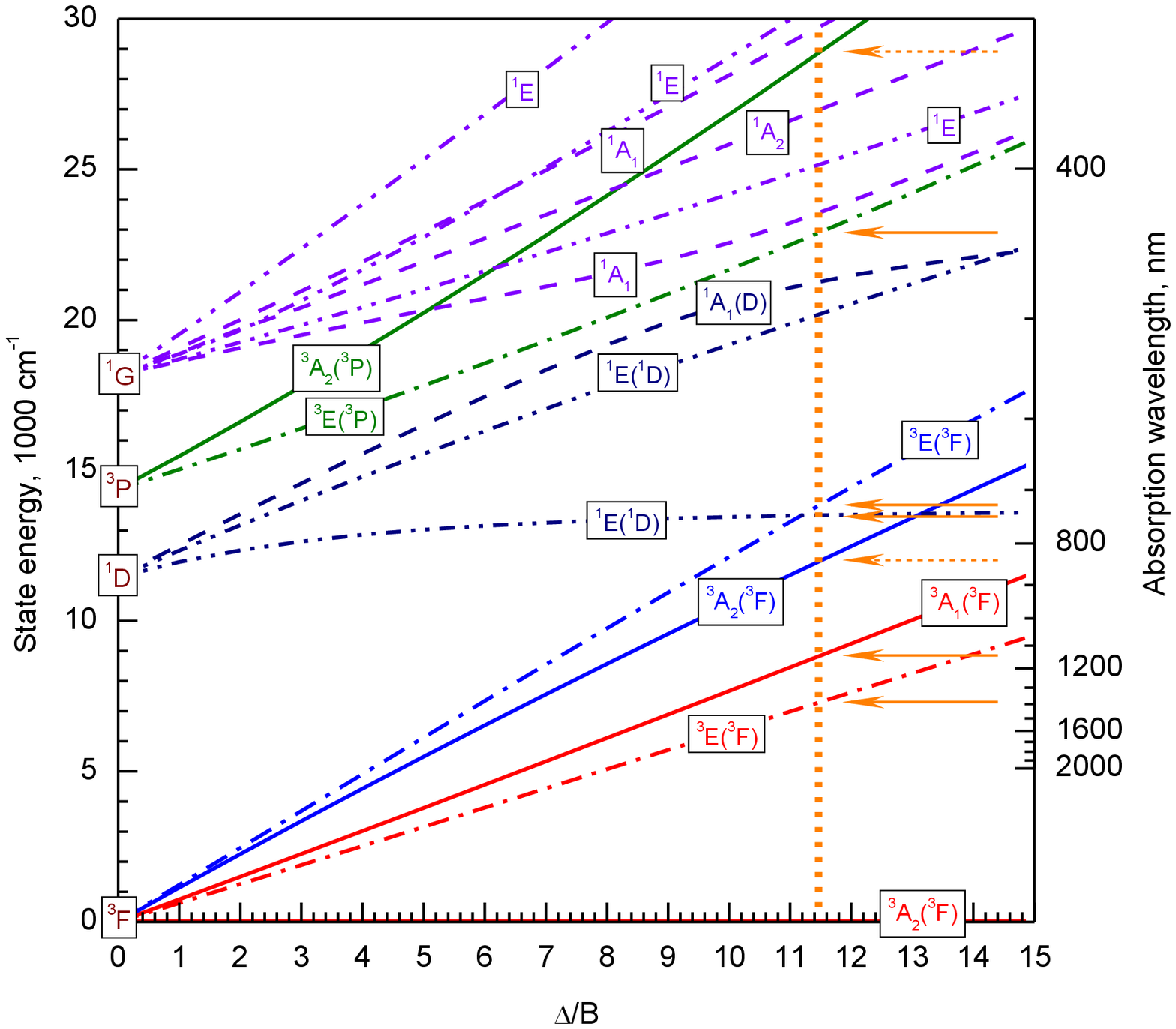}
\caption{Tanabe-Sugano diagram for \Niiip{} ion in \WOiiiTeOiixNi{}{}{}
tellurite glass. Vertical dashed line corresponds to $\Delta/B \approx 11.5$.
Wavelengths of \Symm{D}{3} symmetry allowed and forbidden transitions are marked
by solid and dashed arrows, respectively}
\label{fig:6}
\end{figure}

\begin{figure*}
\includegraphics[scale=0.65,bb=0 -30 725 1005]{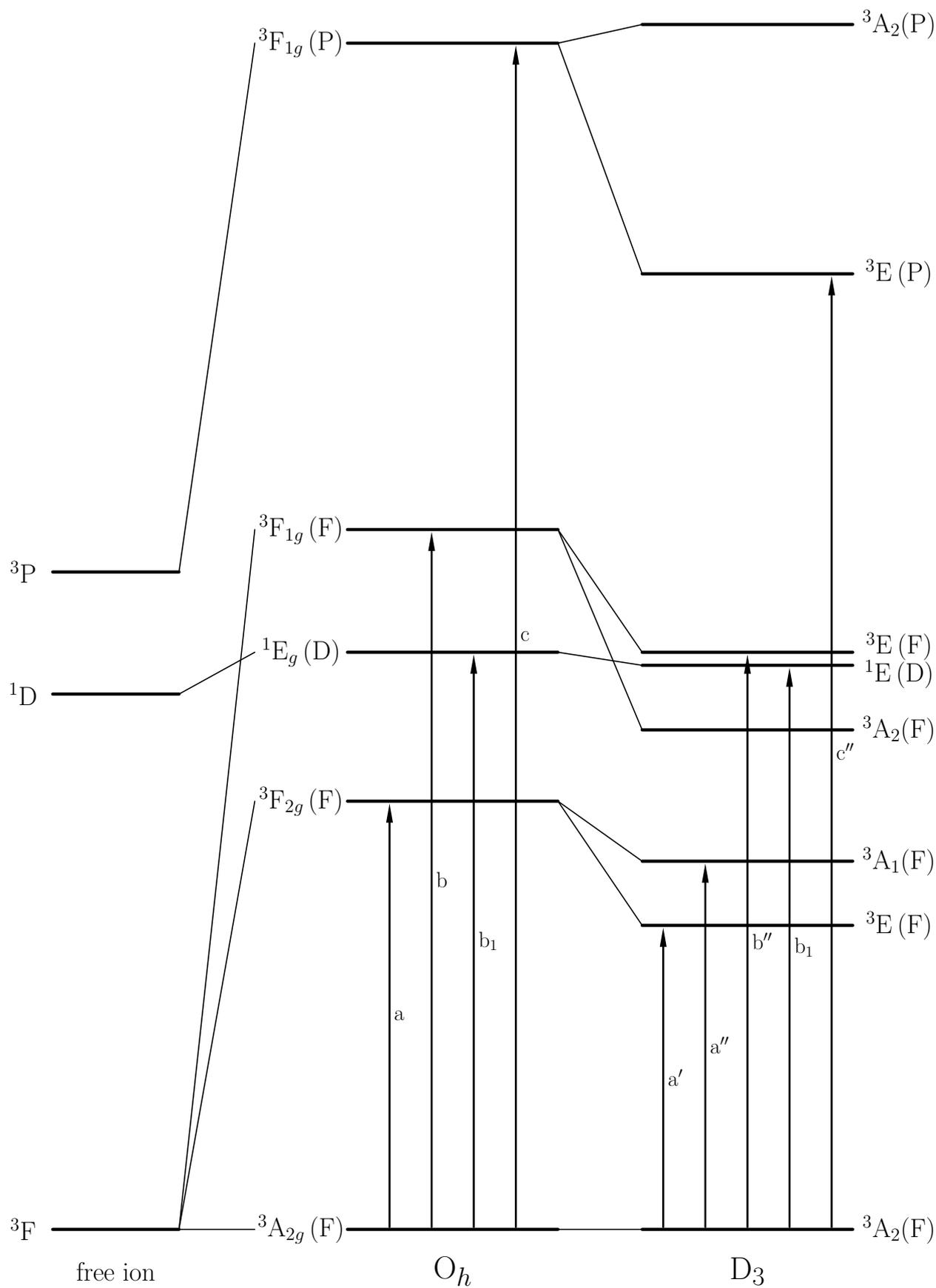}
\caption{Energy levels and main E1-transitions in octahedrally coordinated
\Niiip{} ion in the \WOiiiTeOiix{}{}{} glass network in \dn{8}{} electronic
configuration in the cubic \Symm{O}{h} and trigonal \Symm{D}{3} environment in
comparison with free \Niiip{} ion}
\label{fig:7}
\end{figure*}

Using the geometric parameters obtained in this modeling we calculated
electronic states of \Niiip{} ion in \WOiiiTeOiix{}{}{} glass network
corresponding to \dn{8} electronic configuration of the \Niiip{} ion. The
calculation was performed using \AOMX{} program \cite{AOMX} in angular overlap
model of the ligand field theory \cite{Jorgensen63, Schaeffier70, Hoggard03}.
The model parameters were optimized demanding the best reproduction of
experimental values of the absorption bands wavelengths: the crystal field
parameter $\Delta \approx 10988$~\cmminusi, Racah parameters $B \approx
958$~\cmminusi{} ($\Delta/B \approx 11.5$) and $C \approx 3330$~\cmminusi, the
angular overlap parameters $e_\sigma \approx 3663$\cmminusi{} and $e_\pi \approx
183$~\cmminusi. With these parameters the Tanabe-Sugano diagram \cite{Tanabe54}
was designed for \Niiip{} in the \dn{8} electronic configuration in
\WOiiiTeOiix{}{}{} glass network. The diagram is shown in Fig.~\ref{fig:6} where
the $\Delta/B$ parameter value corresponding to the \Niiip{} absorption bands
experimentally observed in \WOiiiTeOiixNi{}{}{} glasses and the transitions
wavelengths are marked. Fig.~\ref{fig:7} presents the scheme of calculated
levels and main E1 transitions in octahedrally coordinated \Niiip{} ion in the
\dn{8} electronic configuration in \WOiiiTeOiix{}{}{} glass network both in
\Symm{O}{h} cubic and \Symm{D}{3} trigonal environments in comparison with free
\Niiip{} ion \cite{Shenstone54}.

Results of the modeling allow us to interpret the observed absorption spectrum
of \Niiip{} ions in \WOiiiTeOiixNi{}{}{} glass as follows.

In the case of cubic \Symm{O}{h} symmetry of the \Niiip{} ion octahedral
environment three main absorption bands are known to occur (see e.g.
\cite{Galoisy91, Solntsev06, Suzuki08}). The bands correspond to spin-allowed E1
transitions from the ground state, \Term{A}{2g}{3}{F}{}{3}, to the excited
states, \Term{F}{2g}{3}{F}{}{3}{} ((a)~transition in Fig.~\ref{fig:7}, typically
in the 1100 -- 1200~nm range); \Term{F}{1g}{3}{F}{}{3}{} ((b)~transition in
Fig.~\ref{fig:7}, typically in 630 -- 770~nm range); and
\Term{F}{1g}{3}{P}{}{3}{} ((c)~transition in Fig.~\ref{fig:7}, typically in 380
-- 500~nm range).

Besides, weak absorption band caused by spin-forbidden transition from the
\Term{A}{2g}{3}{F}{}{3}{} ground state to the \Term{E}{g}{1}{D}{}{1}{} excited
state is sometimes observed near the (b)~band ((b$_1$)~transition in
Fig.~\ref{fig:7}). Notice to avoid confusion that all these transitions are
parity-forbidden but become slightly allowed owing to vibronic contributions.
Calculation using the above-listed parameters values yields the these
transitions wavelengths for \Niiip{} ion in \Symm{O}{h}{} environment to be
(a)~1000~nm, (b)~600~nm, (c)~350~nm, (b$_1$) 723~nm, in satisfactory agreement
with the typical values for \Symm{O}{h}{} environment given above.

In \Symm{D}{3d} or \Symm{D}{3} trigonally distorted octahedral environment of
the \Niiip{} ion, the ground state turns out to be \Term{A}{2(g)}{3}{F}{}{3}{}
and each of the \Term{F}{1(g)}{3}{F}{}{3}, \Term{F}{2(g)}{3}{F}{}{3}, and
\Term{F}{1(g)}{3}{P}{}{3}{} excited states is split into two states,
\Term{A}{1(g)}{3}{F}{}{3}, \Term{E}{(g)}{3}{F}{}{3}; \Term{A}{2(g)}{3}{F}{}{3},
\Term{E}{(g)}{3}{F}{}{3}; \Term{A}{2(g)}{3}{P}{}{3}, \Term{E}{(g)}{3}{P}{}{3},
respectively. The \Term{E}{(g)}{1}{D}{}{1}{} is not split. In the case of
\Symm{D}{3}{} distortion E1 transitions from the \Term{A}{2}{3}{F}{}{3}{} ground
state to the excited \Term{A}{1)}{3}{F}{}{3}, \Term{E}{}{3}{F}{}{3}{} and
\Term{E}{}{3}{P}{}{3}{} turn out to be both spin- and symmetry-allowed ((a$'$),
(a$''$), (b$''$) and (c$''$) transitions in Fig.~\ref{fig:7}) while transitions
to the \Term{A}{2)}{3}{F}{}{3}{} and \Term{A}{2)}{3}{P}{}{3}{} excited states
are symmetry-forbidden. In the case of \Symm{D}{3d}{} distortion all the E1
transitions from the \Term{A}{2g}{3}{F}{}{3}{} ground state to the
\Term{A}{1g}{3}{F}{}{3}, \Term{E}{g}{3}{F}{}{3}, \Term{E}{g}{3}{P}{}{3},
\Term{A}{2g}{3}{F}{}{3}, and \Term{A}{2g}{3}{P}{}{3}{} excited states, are
symmetry-allowed. The transition from the \Term{A}{2g}{3}{F}{}{3}{} ground state
to the \Term{E}{g}{1}{D}{}{1}{} excited one is still spin-forbidden ((b$_1$)
transition in Fig.~\ref{fig:7}). The remark concerning the vibronic
contributions remains valid.

In our case, (a$'$) and (a$''$) transitions from \Term{A}{2(g)}{3}{F}{}{3}{} to
\Term{A}{1(g)}{3}{F}{}{3}{} and \Term{E}{(g)}{3}{F}{}{3}{} correspond to
the absorption band near 1300~nm; (b$''$) transition from
\Term{A}{2(g)}{3}{F}{}{3}{} to \Term{E}{(g)}{3}{F}{}{3}{} corresponds to the
absorption band near 800~nm; (b$_1$) transition from \Term{A}{2(g)}{3}{F}{}{3}{}
to \Term{E}{(g)}{1}{D}{}{1}{} corresponds to the weakly pronounced band near
740~nm; and (c$''$) transition from \Term{A}{2(g)}{3}{F}{}{3}{} to
\Term{E}{g}{3}{P}{}{3} corresponds to the absorption band in the $< 500$~nm
range.

Thus, in binary tungstate-tellurite glasses, due to strong trigonal distortion
of the octahedral environment of \Niiip{} ions, absorption bands of these ions
turn out to be shifted substantially towards longer wavelengths in comparison
with the absorption pattern typical for nickel-doped crystals. In this case,
obviously, the IR absorption band (near 1320~nm) should be strongly
heterogeneously broadened while the (heterogeneous) broadening of the other
absorption bands should be relatively weak.

\section{Summary}
\label{Summary}
In this study we measured transmission spectra of \WOiiiTeOiixNi{22}{78}{}
tungstate tellurite glass samples containing from 1 to $1.2 \cdot
10^{-2}$~wt.\%~NiO and obtained spectral dependence of \Niiip{} ions extinction
coefficient in this glass in the 450 -- 2700~nm wavelength range. These
measurements together with our computer modeling of \WOiiiTeOiixNi{}{}{} glass
structure suggest \Niiip{} ions in such glasses to occur in strongly trigonally
distorted octahedral sites. This results in corresponding absorption bands
shifted significantly towards longer wavelengths compared to typical
nickel-doped crystals and the absorption intensity being much higher than in
silica- and zirconium fluoride-based glasses. So regarding both width of the
absorption spectral range and the extinction coefficient value, nickel should be
considered among strongly absorbing impurities in the tellurite glasses.

%
%

\end{document}